\documentclass[aps,prd,floatfix,superscriptaddress,twocolumn,showpacs]{revtex4}

\usepackage{epsfig}
\usepackage{amsmath}
\usepackage{graphicx,psfrag}
\usepackage{dcolumn}
\usepackage{bm}
\usepackage{slashed}
\usepackage{epstopdf}

\newcommand{\beq}{\begin{equation}}
\newcommand{\eeq}{\end{equation}}
\newcommand{\bea}{\begin{eqnarray}}
\newcommand{\eea}{\end{eqnarray}}
\newcommand{\hf} {\frac{1}{2}}

\newcommand{\nn}{\nonumber\\}

\newcommand\eqn[1]     {Eq.\,(\ref{#1})}

\newcommand\fig[1]     {Fig.\,{\ref{#1}}}

\def\eq#1{(\ref{#1})}
\def\s0#1#2{\mbox{\small{$ \frac{#1}{#2} $}}}
\def\0#1#2{\frac{#1}{#2}}

\def\tu{{\tilde u}}

\def\mr#1{{\mathrm{#1}}}

\begin{document}

\title{$c$-function and central charge of the sine-Gordon model from the 
non-perturbative renormalization group flow}

\author{V. Bacs\'o}
\affiliation{University of Debrecen, P.O.Box 105, H-4010 Debrecen, Hungary}

\author{N. Defenu}
\affiliation{SISSA, via Bonomea 265, I-34136 Trieste, Italy.}
\affiliation{CNR-IOM DEMOCRITOS Simulation Center, Via Bonomea 265, I-34136 Trieste, Italy.}

\author{A. Trombettoni} 
\affiliation{CNR-IOM DEMOCRITOS Simulation Center, Via Bonomea 265, I-34136 Trieste, Italy.}
\affiliation{SISSA, via Bonomea 265, I-34136 Trieste, Italy.}

\author{I. N\'andori}
\affiliation{MTA-DE Particle Physics Research Group, P.O.Box 51, H-4001 Debrecen, Hungary}
\affiliation{MTA Atomki, P.O. Box 51, H-4001 Debrecen, Hungary} 

\begin{abstract} 
In this paper we study the $c$-function of the sine-Gordon model taking explicitly into account 
the periodicity of the interaction potential. The integration of the $c$-function along trajectories 
of the non-perturbative renormalization group flow gives access to the central charges of the model 
in the fixed points. The results at vanishing frequency $\beta^2$, where the periodicity does not play 
a role, are retrieved and the independence on the cutoff regulator for small frequencies is discussed.
Our findings show that the central charge obtained integrating the trajectories starting from the 
repulsive low-frequencies fixed points ($\beta^2 <8\pi$) to the infrared limit is in good quantitative 
agreement with the expected $\Delta c=1$ result. The behavior of the $c$-function in the other parts of the 
flow diagram is also discussed. Finally, we point out that also including higher harmonics in the 
renormalization group treatment at the level of local potential approximation is not sufficient to give reasonable 
results, even if the periodicity is taken into account. Rather, incorporating the wave-function 
renormalization (i. e. going beyond local potential approximation)  is crucial to get sensible results even when a single frequency is used. 
\end{abstract}

\pacs{11.10.Hi, 05.70.Fh, 64.60.-i, 05.10.Cc}

\maketitle

\section{Introduction}
\label{sec_intro}

Statistical field theory has known an outpouring development in the last decades,
with the systematic improvement of powerful theoretical tools for the study of critical phenomena 
and phase transitions \cite{giuseppe}. 
A key role among these methods is played by the renormalization group (RG) approach: RG 
allowed to treat statistical physics models studying the behavior of 
the transformations bringing the microscopic variables into macroscopic ones \cite{Goldenfeld,Cardy}.
Using RG one can have not only a qualitative picture of the phase diagram and fixed points, 
but also accurate quantitative estimates of critical properties as critical 
exponents and universal quantities - even though very few cases 
are known where RG procedure can be carried out exactly and the method 
itself offers few possibility to obtain exact results.

On the other hand, the scale invariance exhibited by systems at criticality may give rise to invariance 
under the larger group of conformal transformations \cite{polyakov_1970} locally acting 
as scale transformations \cite{di_francesco_1997}.  
The conformal group in $d$ spatial dimensions (for $d\neq2$) 
has a number of independent generators
equal to $\frac{1}{2}(d+1)(d+2)$, while for $d=2$ the conformal
group is infinitely dimensional \cite{di_francesco_1997}. The occurrence and consequences of conformal invariance 
for $2$-dimensional field theories have been deeply investigated and exploited 
to obtain a variety of exact results \cite{di_francesco_1997,giuseppe} and a systematic  
understanding of phase transitions in two dimensions.

A bridge between conformal field theory (CFT) techniques and the RG description of field theories 
is provided in two dimensions by the $c$-theorem. 
Far from fixed points, Zamolodchikov's $c$-theorem \cite{c_theorem} can be used to get information 
on the scale-dependence of the model. In particular the theorem states that it is always possible to 
construct a function of the couplings, the so-called $c$-function, which monotonically decreases when evaluated 
along the trajectory of the RG flow. Furthermore, at the fixed points this function assumes the same value as 
the central charge of the corresponding CFT \cite{giuseppe}.

Although the $c$-theorem is by now a classical result, the determination of the $c$-function is not 
straightforward and its computation far from fixed points is non-trivial 
even for very well known models, so that methods as form factor perturbation theory, truncated conformal space approach 
and conformal perturbation theory has been developed \cite{giuseppe}. 
In $d=2$ an expression of the $c$-function has been obtained 
in the framework of form factor perturbation theory \cite{cardy_prl} for theories away
from criticality and it has been applied to the sinh-Gordon model \cite{fring}. The 
sinh-Gordon model is a massive integrable scalar theory, with no phase transitions. In \cite{fring} one finds $\Delta c=1$ 
for the sinh-Gordon theory. In a recent result \cite{dorey}, the analytical continuation 
of the sinh-Gordon $S$-matrix produces a roaming phenomenon 
exhibiting $\Delta c=1$ and multiple plateaus of the $c$-function.
The analytic continuation $\beta \to i\beta$ of the sinh-Gordon model 
leads to the well-known 
sine-Gordon (SG) model with a periodic self interaction of the form $\cos{(\beta \phi)}$. 

The SG model presents the unique feature
to have a whole line of interacting fixed points with coupling (temperature) dependent critical exponents.
It is in the same universality of the $2$-dimensional Coulomb gas \cite{minnaghen} and 
of the $2$-dimensional $XY$ \cite{kadanoff}, 
thus being one of the most relevant and studied $2$-dimensional model, with applications 
ranging from the study of the Kosterlitz-Thouless transition \cite{kadanoff} to quantum integrability \cite{korepin} 
and bosonization \cite{tsvelik}. In particular, for the SG model an ubiquitous issue is how to 
deal with the issue of the periodicity of the field \cite{coleman}, which unveils and plays a crucial 
role for $\beta \neq 0$. Given the importance of the SG as a paradigmatic $2$-dimensional model, 
the determination of the $c$-function from the non-perturbative RG flow is a challenging goal, 
in particular to clarify the role played by the periodicity of the field for $\beta \neq 0$.

From the RG point of view, the determination of the behavior of the $c$-function is a challenging task 
requiring a general non-perturbative 
knowledge of the RG flow. Recently \cite{c_func}, an expression for the Zamolodchikov's $c$-function has been 
derived for $2$-dimensional models in the Functional RG (FRG) framework \cite{Berges02,Polony04,Delamotte12}: 
resorting to an approximation well established and studied  
in the FRG, the Local Potential Approximation (LPA), an approximated and concretely computable 
RG flow equation for the $c$-function was also written down \cite{c_func}. By using this expression 
known results were recovered for scalar models 
on some special trajectories of the Ising and SG models. For the SG model, having a Lagrangian 
proportional to $\cos{(\beta \phi)}$, the determination and 
the integration of the $c$-function was carried out for $\beta=0$ 
as a massive deformation of the Gaussian fixed point \cite{c_func}. 
Motivated by these results, both for the Ising and SG models and for general $2$-dimensional models, 
it would be highly desirable to have a complete 
description of the $c$-function on general RG trajectories.

In the present paper we present the first numerical calculation of the 
$c$-function on the whole RG flow phase diagram of the SG model. The goal is to determine the behavior of
the $c$-function, and the presence of known results (namely, $\Delta c=1$) helps to assess the validity 
of our approach along the different flows. We also complete the description initiated 
in \cite{c_func} moving to more complex trajectories and showing that these cases are not a 
straightforward generalization of the known results. We finally discuss the dependence of these 
results on the approximation scheme used to compute FRG equations.

\section{Functional Renormalization Group Method}
\label{sec_rg}
In this section, we briefly summarize the FRG approach \cite{Berges02,Polony04,Delamotte12}.
Starting from the usual concepts of Wilson's renormalization group (RG) it is possible to derive
an exact flow equation for the effective action of any quantum field theory.
Tßhis flow equation is commonly written in the form 
\beq
\label{RG}
k \partial_k \Gamma_k [\varphi] = \hf \mathrm{Tr}  
\left[\frac{k \partial_k R_k }{\Gamma^{(2)}_k [\varphi] + R_k} \right],
\eeq 
where $\Gamma_k [\varphi]$ is the effective action and $\Gamma^{(2)}_k [\varphi]$ 
denotes the second functional derivative of the effective action. The trace Tr stands 
for an integration over all the degrees of freedom of the field $\phi$, while 
$R_{k}$ is a regulator function depending on the mode of the field and on the running 
scale $k$. When the running scale goes to zero $k\to 0$ the scale-dependent effective 
action $\Gamma_{k=0}[\phi]$ is the exact effective action of the considered quantum field theory.

Usually equation \eqref{RG} is treated in momentum space, thus the trace stands for a momentum
integration and the regulator $R_{k}$ is a smooth function which freezes all the modes
with momentum smaller than the scale $k$.

The exact FRG equation \eq{RG} stands for functionals, thus it is handled 
by truncations. Truncated RG flows depend on the choice of the regulator function $R_k$, i.e. on 
the renormalization scheme. Regulator functions have already been discussed in the literature by 
introducing their dimensionless form
\beq
R_k( p) = p^2 r(y),
\hskip 0.5cm
y=p^2/k^2,
\eeq
where $r(y)$ is dimensionless. Various types of regulator functions can be chosen (an archetype of 
regulator functions \cite{css} has been shown to take the forms of the regulators used so far by
setting its parameters). In this work we are going to consider the following regulators 
\begin{subequations}
\begin{equation}
\label{Morris_Family}
r_{\mr{pow}}(y)=\frac{1}{y^b},
\end{equation}
\begin{equation}
\label{Litim_Family}
r_{\mr{opt}}(y)=\left(\frac{1}{y} -1\right) \theta(1-y),
\end{equation}
\end{subequations}
where the first is known as the power-law type \cite{Mo1994} and the second one is the 
Litim (or optimized) \cite{opt_rg} type regulator. Let us note that the so called mass cutoff 
regulator, which is used in \cite{c_func}, is identical to $r_{\mr{pow}}(y)$ with $b=1$.

One of the commonly used systematic approximations is the truncated derivative 
expansion, where the action is expanded in powers of the derivative of the field \cite{Berges02}:
\beq
\Gamma_k [\varphi] = \int_x \left[V_k(\varphi) 
+ Z_k(\varphi) \hf (\partial_{\mu} \varphi)^2 + \cdots \right].  
\nonumber
\eeq 
In LPA higher derivative terms are neglected and 
the wave-function renormalization is set equal to constant, i.e. $Z_k \equiv 1$. In this
case \eq{RG} reduces to the partial differential equation for the dimensionless blocked 
potential ($\tilde V_k = k^{-2} V_k$) which has the following form in 2 dimensions 
\bea
\label{lpa}
(2+k\partial_k) {\tilde V}_k(\varphi) = - \frac{1}{4\pi} \int_0^\infty dy 
\frac{y^2 \frac{dr}{dy}}{(1+r)y + {\tilde V''}_k(\varphi)}.
\eea

\section{The $c$-function in the Framework of Functional Renormalization Group}
\label{sec_rg_c}

An expression for the $c$-function in FRG was recently developed in \cite{c_func}. 
In this section we are going to give 
the guidelines of this derivation, reviewing the main results used in the next sections.

Let us start considering an effective action $\Gamma[\phi,g]$ for a single field $\phi$ in curved 
space, with metric $g_{\mu\nu}$. We can study the behavior of this effective action under 
transformation of the field and the metric:
\begin{eqnarray}
\label{Conf_trans1}
\phi \rightarrow &e^{d_{\phi}\tau} \phi \\
\label{Conf_trans2}
g_{\mu\nu} \rightarrow &e^{2\tau} g_{\mu\nu}
\end{eqnarray}
where $d_{\phi}$ is the conformal weight of the field ($d_{\phi}=-\frac{d-2+\eta}{2}$ for a scalar field) while 
the background metric $g_{\mu\nu}$ has always conformal weight $2$. From the requirement that the 
effective action must be invariant under the Weyl transformation \eqref{Conf_trans1}-\eqref{Conf_trans2}, 
we obtain the following expression for a conformal field theory (CFT) in curved space \cite{c_func},
\begin{equation}
\label{EAA_split}
\Gamma[\phi,g]=S_{CFT}[\phi,g]+cS_{P}[g].
\end{equation}
$S_{CFT}[\phi,g]$ is the curved space generalization of the standard CFT action, which is recovered 
in the flat space case $g_{\mu\nu}=\delta_{\mu\nu}$, $c$ is the central charge of our theory and 
$S_{P}[g]$ is the Polyakov action term which is necessary to maintain the Weyl invariance of the 
effective average action in curved space. 

To obtain FRG equations one has to add an infra-red (IR) cutoff term $\Delta S_{k}[\phi,g]$ 
to the ultra-violet (UV) action of the theory. This is a mass term which depends both on the momentum of the 
excitations and on a cutoff scale $k$.
\begin{equation}
\label{cutoff_action}
\Delta S_{k}[\phi,g]=\frac{1}{2}\int{d^{2}x\sqrt{g} \phi(x) R_{k}(\Delta)\phi(x)},
\end{equation}
where $\Delta$ is the spatial Laplacian operator. The effect 
is to freeze the excitation of momentum $q \ll k$, but leaving the excitation at $q>k$ almost 
untouched. The result of this modification of the UV action is to generate, after integrating over the field 
variable, a scale-dependent effective action $\Gamma_{k}[\phi,g]$ which describes our theory at scale $k$. 
When the scale $k$ is sent to zero the cutoff term in the UV action vanishes and the $\Gamma_{k}[\phi,g]$ 
is the exact effective average action of the theory.

The generalization of \eqref{EAA_split} in presence of the cutoff terms is
\begin{equation}
\label{EAA_split_scaling}
\Gamma_{k}[\phi,g]=S_{k}[\phi,g]+c_{k}S_{P}[g]+\cdots,
\end{equation}
where $c_{k}$ is now the scale-dependent $c$-function and the dots stands from some geometrical terms 
which do not depend on the field. We should now consider the case of a flat metric with a dilaton
background $g_{\mu\nu}=e^{2\tau}\delta_{\mu\nu}$. Using the standard path integral formalism for the effective
action we can write
\begin{equation}
 \label{Path_Integral_Gamma_Definition}
 \begin{split}
&e^{-\Gamma_{k}[\phi,e^{2\tau}\delta]}=e^{-S_{k}[\phi,e^{2\tau}\delta]-c_{k}S_{P}[e^{2\tau}\delta]}=\\
&\int{\mathcal{D}\chi_{d.b.} e^{-S_{UV}[\phi+\chi,e^{2\tau}\delta]-c_{UV}S_{P}[e^{2\tau}\delta]-\Delta S_{k}[\chi,e^{2\tau}\delta]}}
\end{split}
\end{equation}
where $S_{UV}[\phi,g]$ is some UV action, $c_{UV}$ is the value of the $c$-function in the UV (which can
be equal to the central charge of some CFT if we are starting the flow from a conformal invariant theory) and
$\chi$ is the fluctuation field. The notation $\mathcal{D}\chi_{d.b.}$ stands for an integration over
the fluctuation field $\chi$ in the curved space of the dilaton background \cite{c_func}. We can further manipulate latter expression moving $c_{UV}$ on the l.h.s
\begin{equation}
 \label{Path_Integral_c_Definition}
 \begin{split}
&e^{-S_{k}[\phi,e^{2\tau}\delta]+(c_{UV}-c_{k})S_{P}[e^{2\tau}\delta]}=\\
&\int{\mathcal{D}\chi_{d.b.} e^{-S_{UV}[\phi+\chi,e^{2\tau}\delta]-\Delta S_{k}[\chi,e^{2\tau}\delta]}}.
\end{split}
\end{equation}
The Polyakov action in the dilaton background case assumes the form
\begin{equation}
 S_{P}[g]=-\frac{1}{24}\int{\tau\Delta\tau},
\end{equation}
where $\tau$ is the dilaton field, $\Delta$ is the laplacian operator and the integral is over
an implicit spatial variable. Substituting latter expression into \eqref{Path_Integral_c_Definition}
 we obtain
\begin{equation}
 \label{Path_Integral_c_Definition2}
 \begin{split}
&e^{-S_{k}[\phi,e^{2\tau}\delta]-\frac{(c_{UV}-c_{k})}{24}\int \tau \Delta \tau} =\\
&\int{\mathcal{D}\chi_{d.b.} e^{-S_{UV}[\phi+\chi,e^{2\tau}\delta]-\Delta S_{k}[\chi,e^{2\tau}\delta]}}.
\end{split}
\end{equation}
In order to recover the usual flat metric integration we have 
to pursue a Weyl transformation \eqref{Conf_trans1} for the fields $\phi$
and $\chi$
\begin{equation}
 \label{Path_Integral_c_Definition_Weyled}
 \begin{split}
&e^{-S_{k}[e^{d_{\phi}\tau}\phi,e^{2\tau}\delta]-\frac{(c_{UV}-c_{k})}{24}\int \tau \Delta \tau} =\\
&\int{\mathcal{D}\chi e^{-S_{UV}[e^{d_{\phi}\tau}(\phi+\chi),e^{2\tau}\delta]-\Delta S_{k}[e^{d_{\phi}\tau}\chi,e^{2\tau}\delta]}},
\end{split}
\end{equation}
and now the integration measure is in flat space.

Finally deriving previous expression with respect to the logarithm of the
FRG scale we obtain that the flow of the $c$-function $\partial_{t}c_{k}$ can be
extracted from the flow of the cutoff action \eqref{cutoff_action} by taking the
coefficient of the $\int \tau \Delta \tau$ term,
\begin{equation}
 \label{Exact_c_function}
 k\partial_{k}c_{k}=24\pi \langle k\tilde{\partial}_{k}\Delta S_{k}[e^{d_{\phi}\tau}\chi,e^{2\tau}\delta]\rangle\Bigl{|}_{\int{\tau\Delta\tau}},
\end{equation}
which after some manipulation becomes \cite{c_func}
\begin{equation}
 \label{Exact_c_function}
 k\partial_{k}c_{k}=-12\pi k\tilde{\partial}_{k}G_{k}[\tau]\Bigl{|}_{\int{\tau\Delta\tau}}.
\end{equation}
Eq. \eqref{Exact_c_function} 
shows that the $c$-function flow is proportional to the coefficient of the $\int{\tau\Delta\tau}$
term in the expansion of the propagator flow $k\tilde{\partial}_{k}G_{k}[\tau]$, also this flow has to
be computed taking into account only the $k$ dependence of the regulator function, i.e.
\begin{equation}
k\tilde{\partial}_{k}=k\partial_{k}R_{k}\frac{\partial}{\partial R_{k}}.
\end{equation}
This equation describes the exact flow of the $c$-function into the FRG framework. 

Since it is not in general possible to solve exactly equation \eqref{RG} and also equation \eqref{Exact_c_function}
needs to be projected into a simplified theory space to be computed numerically. 
In Ref. \cite{c_func}  an explicit expression for the flow equation of the $c$-function in the LPA scheme
has been derived  with the mass cutoff
\bea
\label{c_function}
k\partial_k c_k = \frac{[k\partial_k \tilde V''_k(\varphi_{0,k})]^2}{[1+ \tilde V''_k(\varphi_{0,k})]^3},
\eea
with the dimensionless blocked potential $\tilde V_k(\varphi)$ which is evaluated at its 
running minimum $\varphi=\varphi_{0,k}$ (i.e. the solution of $\tilde V'_k(\varphi) = 0$).
 We observe that an explicit expression for the $c$-function beyond LPA is not available in
literature.

It should be noticed that, while \eqref{lpa} is valid for any regulator (cutoff) function, the expression 
for the $c$-function \eqref{c_function} has been obtained by using the mass cutoff, i.e. \eqref{Morris_Family} 
with $b=1$. Other cutoff choices proved to be apparently very difficult to investigate. 
In the following, we will argue that while the expression \eqref{c_function} 
is sufficient to obtain a qualitative (and almost quantitative) 
picture of the $c$-function phase diagram the usage of other regulator functions is 
necessary to achieve full consistency.  Where it is possible we will check the cutoff dependence of
our numerical results.

\section{RG study of the sine-Gordon model}
\label{sec_sg}
The SG scalar field theory is defined by the Euclidean action for $d=2$
\beq
\label{sg}
\Gamma_{k}[\phi] = \int d^2 x \left[\hf (\partial_\mu\varphi_x)^2 - u\cos(\beta \varphi_x)\right],
\eeq
where $\beta$ and $u$ are the dimensional couplings. Since we are interested in the FRG 
study of the SG model which is periodic in the field variable, the symmetry of the action under the 
transformation \cite{nandori_sg}
\bea
\label{pertr}
\varphi(x) \to \varphi(x) + {\cal A}
\eea
is to be preserved by the blocking and the potential $\tilde V_k(\varphi)$ must be periodic with 
period length ${\cal A}$. It is actually obvious that the blocking, i.e. the transformation given by 
replacing the derivative with respect to the scale $k$ by a finite difference in \eq{lpa} preserves 
the periodicity of the potential \cite{nandori_sg,schemes}.

\subsection{The FRG equation for the SG model for scale-independent frequency.}
In LPA one should look for the solution of \eq{lpa} among the periodic function 
which requires the use of a Fourier expansion. When considering a single Fourier mode, the 
scale-dependent blocked potential reads
\beq
\label{lpa_sg}
\tilde V_k(\varphi) =  - \tilde u_k \cos(\beta\varphi),
\eeq
where $\beta$ is scale-independent.

In the mass cutoff case, i.e. the power law regulator \eq{Morris_Family} with $b=1$, one can 
derive \cite{nandori_msg} the flow equation for the Fourier amplitude of \eq{lpa_sg} from \eqn{lpa}:
\bea
\label{wh}
(2+ k \partial_k) \tilde u_k = \frac{1}{2\pi \beta^2 \tilde u_k} 
\left[1 - \sqrt{1-\beta^4 \tilde u^2_k} \right]
\eea
(see Eq. (21) of \cite{nandori_msg} for vanishing mass). Similarly, using the optimized 
regulator \eq{Litim_Family} gives 
%
\bea
\label{litim}
(2 + k \partial_k) \tilde u_k = \frac{1}{2\pi \beta^2 \tilde u_k} 
\left[\frac{1}{\sqrt{1-\beta^4 \tilde u^2_k}} -1 \right].
\eea
%

\subsection{The FRG equation for the SG model for scale-dependent frequency.}
A very simple, but still sensible, modification to ansatz \eqref{sg} is the inclusion of a scale 
dependent frequency, which, in order to explicitly preserve periodicity, should be rather 
considered as a running wave-function renormalization. The ansatz then becomes
\beq
\label{eaans}
\Gamma_k = \int d^2x \left[\frac{1}{2} z_k (\partial_\mu\varphi_x)^2+V_k(\varphi_x)\right],
\eeq
where the local potential contains a single Fourier mode 
\beq
\label{z_lpa_sg}
V_k(\varphi) =  - u_k \cos(\varphi),
\eeq
and the following notation has been introduced
\beq
\label{identifications}
z_{k} \equiv 1/\beta_{k}^2
\eeq
via the rescaling of the field $\varphi \to \varphi/\beta_{k}$ in \eq{sg}, where $z_k$ plays the
role of a field-independent wave-function renormalization. Then \eqn{RG} leads to the evolution 
equations
\bea
\label{ea_v}
& k\partial_k V_k = \hf\int_p{\cal D}_kk\partial_k R_k,\\
\label{ea_z}
& k\partial_k z_k =
{\cal P}_0 V'''^2_k\int_p{\cal D}_k^2k\partial_k R_k\left(
\frac{\partial^2{\cal D}_k}{\partial p^2\partial p^2}p^2
+\frac{\partial{\cal D}_k}{\partial p^2}
\right)
\eea
with ${\cal D}_k=1/(z_k p^2+R_k+V''_k)$ and ${\cal P}_0=(2\pi)^{-1}\int_0^{2\pi} d\varphi$ is the 
projection onto the field-independent subspace. The scale $k$ covers the momentum interval 
from the UV cutoff $\Lambda$ to zero.  It is important to stress that Eqs.~\eq{ea_v}-\eq{ea_z} are
directly obtained using power-law cutoff functions. One may expect that these equations continue to be 
valid for a general cutoff provided that $R_k \to z_k R_k$ \cite{Berges02}. This substitution has been 
tested for $O(N)$ models, but its validity has been not yet discussed in the literature for the SG model.

Inserting the ansatz \eq{z_lpa_sg} into Eqs.~\eq{ea_v} 
and \eq{ea_z} the RG flow equations for the coupling constants can be written as \cite{SG_Trun}
\bea
\label{general_ea_u}
k\partial_k u_k &=&
\frac1{2\pi} \int_p ~\frac{k\partial_k R_k}{u_k}\left(\frac{P_{k}}{\sqrt{P_k^2-u_k^2}}-1\right),\\
\label{general_ea_z}
k\partial_k z_k &=& \frac1{2\pi}\int_p k\partial_k R_k
\biggl(\frac{u_k^2 p^2 (\partial_{p^2}P_k)^2(4P_k^2+u_k^2)}{4(P_k^2-u_k^2)^{7/2}}\nn
&&-\frac{u_k^2P_k(\partial_{p^2}P_k+p^2\partial_{p^2}^2P_k)}{2(P_k^2-u_k^2)^{5/2}} \biggr)
\eea
with $P_k = z_k p^2 + R_k$. 
In general, the momentum integrals have to be performed 
numerically, however in some cases analytical results are available. Indeed, by 
using the power-law regulator \eqref{Morris_Family} with $b=1$, the momentum integrals 
can be performed \cite{nandori_msg} and the RG flow equations read as 
\bea
\label{single_b1_exact}
(2+k\partial_k)\tu_k &=& \frac{1}{2\pi z_k \tu_k}  \left[1 -  \sqrt{1 - \tu_k^2} \right] \nn
k\partial_k z_k &=& -\frac{1}{24\pi} \frac{\tu_k^2}{[1 - \tu_k^2]^\frac{3}{2}}
\eea
with the dimensionless coupling $\tu = k^{-2} u$. By using the replacements
\begin{subequations}
\begin{equation}
\label{Lpatransform}
z_{k} \to 1/\beta_{k}^2, 
\end{equation}
\begin{equation}
\label{Lpatransform2}
\tilde{u}_k \to \beta_{k}^2 \tilde{u}_k, 
\end{equation}
\end{subequations}
and keeping the frequency scale-independent ($\partial_k z_k =0$ i.e. $\partial_k \beta^2_k =0$) 
one recovers the corresponding LPA Eq. \eq{wh}.

\section{$c$-function of the sine-Gordon model for $\beta=0$}
\label{sec_c_func_sg}
In this section we discuss the case $\beta=0$. We start by summarizing the results obtained for 
the $c$-function of the SG model in \cite{c_func}. 
The ansatz considered in \cite{c_func} is 
\bea
\label{sg_scale}
\tilde V_k(\varphi) = - \frac{\tilde m_k^2}{\beta_k^2} \left(\cos(\beta_k \varphi) -1\right),
\eea
where the frequency $\beta_k$ is assumed to be scale-dependent. If one directly substitutes
\eq{sg_scale} into the RG Eq. \eq{lpa}, 
then the l.h.s. of \eq{lpa} generates non-periodic terms due to the scale-dependence of $\beta_k$. Thus,
the periodicity of the model is not preserved and one can use the Taylor expansion of the 
original periodic model 
\bea
\label{sg_poly}
\tilde V_k(\varphi) \approx \hf \tilde m_k^2 \varphi^2 - \frac{1}{4!} (\tilde m_k^2 \beta_k^2) \varphi^4.
\eea
In this case, \eq{sg_scale} is treated as a truncated Ising model and the RG equations 
for the coupling constants read as
\bea
\label{poly_flow_1}
k \partial_k \tilde m_k^2  &=& \frac{\tilde m_k^2  [\beta_k^2  -8\pi(1+\tilde m_k^2)]}{4\pi (1+\tilde m_k^2)} \\
\label{poly_flow_2}
k \partial_k \beta_k^2 &=&  - \frac{1}{4\pi} \frac{(1+4 \tilde m_k^2) \beta_k^4}{(1+\tilde m_k^2)^2}.
\eea
The disadvantage of the scale-dependent frequency is that the periodicity of the model is violated 
changing the known phase structure of the SG model. However, the authors of \cite{c_func} were 
interested in the massive deformation of the Gaussian fixed point which is at $\beta = 0$ and $\tilde u =0$, 
so one has to take the limit $\beta \to 0$ where the Taylor expansion represents a good approximation for 
the original SG model. Indeed, in the limit $\beta \to 0$, the RG Eqs. \eq{poly_flow_1}, \eq{poly_flow_2} 
reduce to
\bea
\label{poly_flow_m}
k \partial_k \tilde m_k^2  &\approx& \frac{\tilde m_k^2  [\beta_k^2  -8\pi(1+\tilde m_k^2)]}{4\pi (1+\tilde m_k^2)} 
\approx -2 \tilde m_k^2\\
\label{poly_flow_beta}
k \partial_k \beta_k^2 &\approx& 0.
\eea
Similar flow equations for the couplings $\tilde m_k^2$ and $\beta_k$ were given in \cite{c_func}. The 
solution for the $c$-function based on \eq{sg_scale} is in agreement with the known exact result, i.e. at the 
Gaussian UV fixed point $c_{\mr{UV}} =1$ and in the IR limit $c_{\mr{IR}} =0$, thus the exact result in case 
of the massive deformation of the Gaussian fixed point is $\Delta c =1$ ($\Delta c = c_{UV}-c_{IR}$). The 
numerical solution \cite{c_func} gives $\Delta c = 0.998$ in almost perfect agreement with the 
exact result. 

Although the numerical result obtained for the $c$-function in \cite{c_func} is more than satisfactory, due to 
the Taylor expansion, the SG theory is considered as an Ising-type model. Thus, the RG study of the $c$-function 
starting from the Gaussian fixed point in the Taylor expanded SG model is essentially the same as that of the 
deformation of the Ising Gaussian fixed point. So, it does not represent an independent check of \eq{c_function}. 
Indeed, inserting \eq{poly_flow_m} into \eq{c_function} using the ansatz \eq{sg_scale} one finds
\bea
\label{c_func_ising}
k\partial_k c_k = \frac{4\tilde m_k^4}{[1+ \tilde m_k^2]^3}
\eea
which is identical to Eq. (5.3) of \cite{c_func} (with $a=1$) obtained for the massive deformation of the Gaussian 
fixed point in the Ising model and it can be also derived from Eq. (5.19) of \cite{c_func} 
in the limit of $\beta^2 \to 0$.

Therefore, it is a relevant question whether one can reproduce the numerical results obtained for the $c$-function 
(with the same accuracy) if the SG model is treated with scale-independent frequency \eq{lpa_sg}, or beyond 
LPA, by the rescaling of the field \eq{z_lpa_sg}. Also ref. \cite{c_func} treats only massive deformations of
non interacting UV fixed points, then on such trajectories only the mass coupling is running. Nevertheless the c-theorem should hold 
on all trajectories, even when more couplings are present. Our aim is to demonstrate that the derivation of 
\cite{c_func} is valid even in these more general cases, but, due to truncation approach, the approximated 
FRG phase diagram does not fulfill the requirements of the c-theorem exactly and, therefore, only approximated results are 
possible.

\section{$c$-function of the sine-Gordon model on the whole flow diagram}
\label{sec_c_func_sg2}
In this section we study the $c$-function of the SG model on the whole phase diagram, studying both the scale 
independent wave-function renormalization and the treatment with the running frequency.

\subsection{Scale-independent frequency case}
The definition for the SG model used in this work, i.e. \eq{sg}, differs from \eq{sg_scale} because the 
frequency parameter is assumed to be scale-independent in LPA. The running of $\beta$ can only be 
achieved beyond LPA by incorporating a wave-function renormalization and using a rescaling of the 
field variable which gives $z_{k} = 1/\beta_{k}^2$. 

\begin{figure}[ht] 
\begin{center} 
\includegraphics[width=0.48\textwidth]{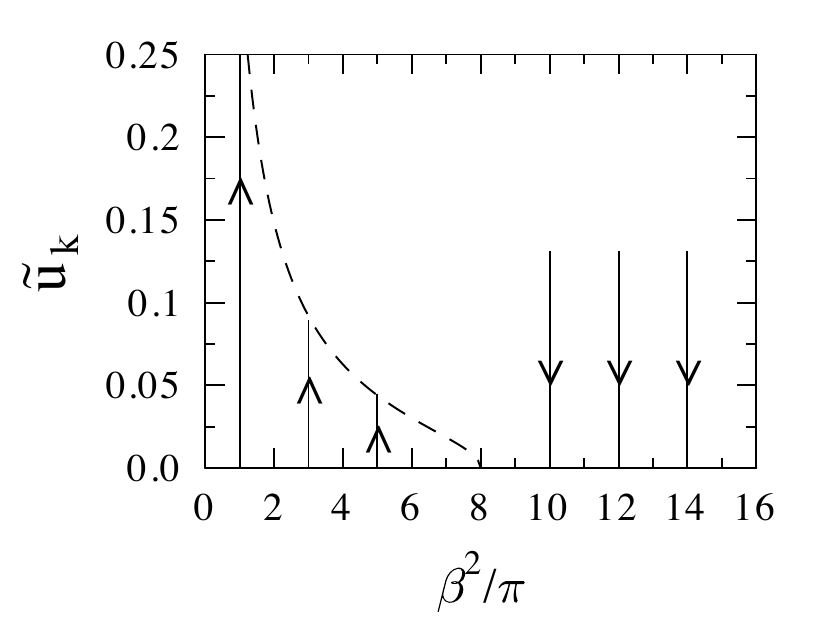}
\caption{
\label{fig1}
 The figure shows the phase structure of the SG model obtained by the FRG equation using Litim's regulator 
in the scale-independent frequency case. The two phases are separated by $\beta_c^2 =8\pi$. The dashed 
line shows the line of IR fixed points of the broken phase.} 
\end{center}
\end{figure}

Let us first discuss the results of LPA. Equations \eq{wh} and \eq{litim} have the same qualitative 
solution.  In \fig{fig1} we show the phase structure obtained by solving \eq{litim}. 
%
%
%

%
%
\begin{figure}[ht] 
\begin{center} 
\includegraphics[width=0.48\textwidth]{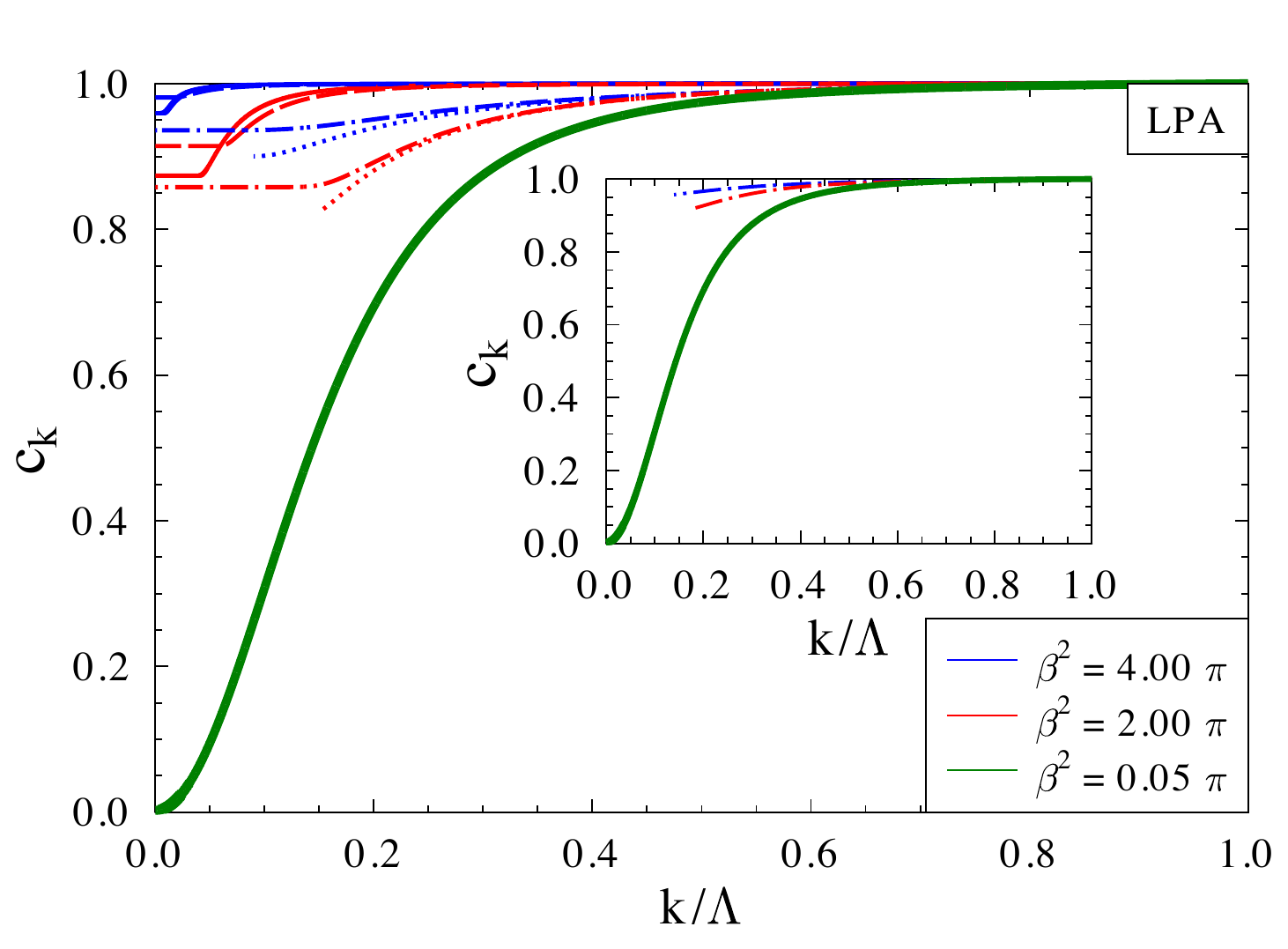}
\caption{
\label{fig1b}
Running of the $c$-function obtained in the scale-independent frequency case by solving 
\eq{wh} (dotted lines - mass cutoff) and \eq{litim} (dot-dashed lines - Litim cutoff) combined with \eq{c_function} for the 
SG model is plotted for various values of the frequency $\beta^2$. From top to bottom it is $\beta^{2}/\pi=4,2,0.005$. Due to the poor 
convergence properties of \eq{wh}, where the mass cutoff was used, the RG flow stops 
at some finite momentum scale and the deep IR value of the $c$-function cannot be reached 
(dotted lines). The use of the Litim cutoff \eq{litim} (dot-dashed lines) can produce us the IR 
constant for the $c$-function. However, for very small value of $\beta^2$ the low-frequency 
approximation is the best choice, i.e. one has to solve \eq{c_func_low_freq_sg} (green line): results from different cutoff functions are indistinguishable. 
The solid lines represent the results obtained with power law cutoff ($b=2$), while the dashed lines are
the results with the exponential cutoff. 
The inset shows the results for an enlarged theory space where higher harmonics are included 
in \eqref{z_lpa_sg} (dot-dashed lines - Litim cutoff).}
\end{center}
\end{figure}
The RG trajectories are straight lines because in LPA the frequency parameter of \eq{sg} is scale 
independent. Above (below) the critical frequency $\beta^2_c = 8\pi$, the line of IR fixed points is at 
$\tilde u_{\mr{IR}} = 0$ ($\tilde u_{\mr{IR}} \neq 0$). For $\beta^2 < 8\pi$ the IR value for the Fourier 
amplitude depends on the particular value of $\beta^2$  thus, one finds different IR effective theories, 
i.e. the corresponding CFT depends on the frequency too.

The scaling for the $c$-function is the one expected from the c-theorem. It is a decreasing function of 
the scale $k$ which is constant in the UV and IR limits, see \fig{fig1}. Due to the approximation of scale 
independent frequency $\beta$, here, the IR value of the $c$-function depends on the particular initial 
condition for $\beta^2$. Then when we start at the Gaussian fixed points line ($c=1$), in the symmetric 
phase, the flow evolves towards an IR fixed point, but at this approximation level, we have different IR 
fixed points which are all at different $\tilde{u}$ values and consequently the $\Delta c$ values differ from 
the exact one. The exact result $\Delta c \simeq 1$ is obtained only in the $\beta\to 0$ limit.

We notice that Eq. \eqref{wh}, where the mass cutoff was used, has very poor convergence properties and
the flow, obtained from them, stops at some finite scale, thus the deep IR values of the $c$-function 
cannot be reached (dashed lines in  Fig. \ref{fig1b}). 

The use of the Litim cutoff RG Eq. \eqref{litim} improves the convergence of the RG flow but the IR results 
for the $c$-function are very far from the expected $\Delta c=1$, which can be recovered only in the 
vanishing frequency limit. Also the inclusion of higher harmonics in \eqref{z_lpa_sg} ( inset in 
 Fig.\ref{fig1b}) does not improve this result. 

It should be noted that Eq. \eqref{c_function} is strictly valid only in the mass cutoff case, however in 
Fig. \ref{fig1b} we used Eq. \eqref{c_function} even in the optimized cutoff case. This inconsistence cannot
be regarded as the cause for the unsatisfactory results obtained in the large $\beta$ cases, indeed we
expect very small dependence of the flow trajectories upon the cutoff choice.

This small dependence on the regulator is evident from the comparison of the mass and Litim 
regulator results of trajectories for the $c$-function in  Fig. \ref{fig1b}, which are very similar, at least in 
the region where no convergence problems are found. This similarity justifies the use of the mass cutoff 
result \eqref{c_function} with RG flow Eqs. \eqref{litim} obtained by the optimized (Litim) regulator. 

 We also computed the $c$-function flow for other cutoff functions, namely the power-law $b=2$ (solid lines in Fig. \ref{fig1b})
 and the exponential one (dashed lines). Apart from the mass cutoff, all the others converge to the IR fixed point.
 The conclusion is that there is not a pronounced dependence of the findings on the cutoff schemes and that 
 the constant frequency case in not sufficient to recover the correct behavior for the $c$-function.

We observe that  the lack of convergence observed in mass cutoff case is not present in the small frequency limit analyzed in \cite{c_func}. Indeed, expanding
flow equations \eq{wh} and \eq{litim} we get
\bea
\label{low_freq_rg}
k \partial_k \tilde u_k \approx -2 \tilde u_k + \frac{\tilde u_k \beta^2}{4\pi} \approx -2 \tilde u_k,
\eea
which is valid for vanishing frequency and it is independent of the particular choice of the regulator function, 
i.e. it is the same for the mass and Litim cutoffs. Substituting  \eq{low_freq_rg} into \eq{c_function} using 
\eq{lpa_sg} the following equation is obtained for the $c$-function of the SG model:
\bea
\label{c_func_low_freq_sg}
k\partial_k c_k = \frac{\left( k\partial_k \tilde u_k \beta^2 \right)^2}
{\left(1+ \tilde u_k \beta^2\right)^3} \approx
\frac{\left(-2 \tilde u_k \beta^2\right)^2}{\left(1+ \tilde u_k \beta^2\right)^3}  \equiv  
\frac{4\tilde m_k^4}{[1+ \tilde m_k^2]^3}
\eea
where the identification $\tilde m_k^2 = \tilde u_k \beta^2$ is used. The scale dependence of the $c$-function 
in that case is identical to the massive deformation of the Gaussian fixed point and the corresponding RG 
trajectory is indicated by the green line in \fig{fig2}. 

It is important to note that for finite frequencies $\beta^2 \neq 0$ the Taylor expanded potential \eq{sg_poly} 
cannot be used to determine the $c$-function since it violates the periodicity of the model. In this case only 
Eqs. \eq{wh} or \eq{litim} can produce reliable results. 

In order to improve the LPA result for the $c$-function of the SG model without violating the periodicity of the 
model one has to incorporate a scale-dependent frequency, i.e. a wave-function renormalization  (we refer to this approximation
as $z+$LPA), as it is discussed in the next subsection.

\subsection{The scale-dependent wave function renormalization }
The inclusion of the running wave-function renormalization changes the whole picture of the SG phase 
diagram, with all the $\tilde{u} \neq 0$ fixed points collapsing into a single ($\beta_{k}=0,\tu=1$) fixed point, 
as it is expected from the exact CFT solution. 
%
%
\begin{figure}[ht] 
\begin{center} 
\includegraphics[width=0.48\textwidth]{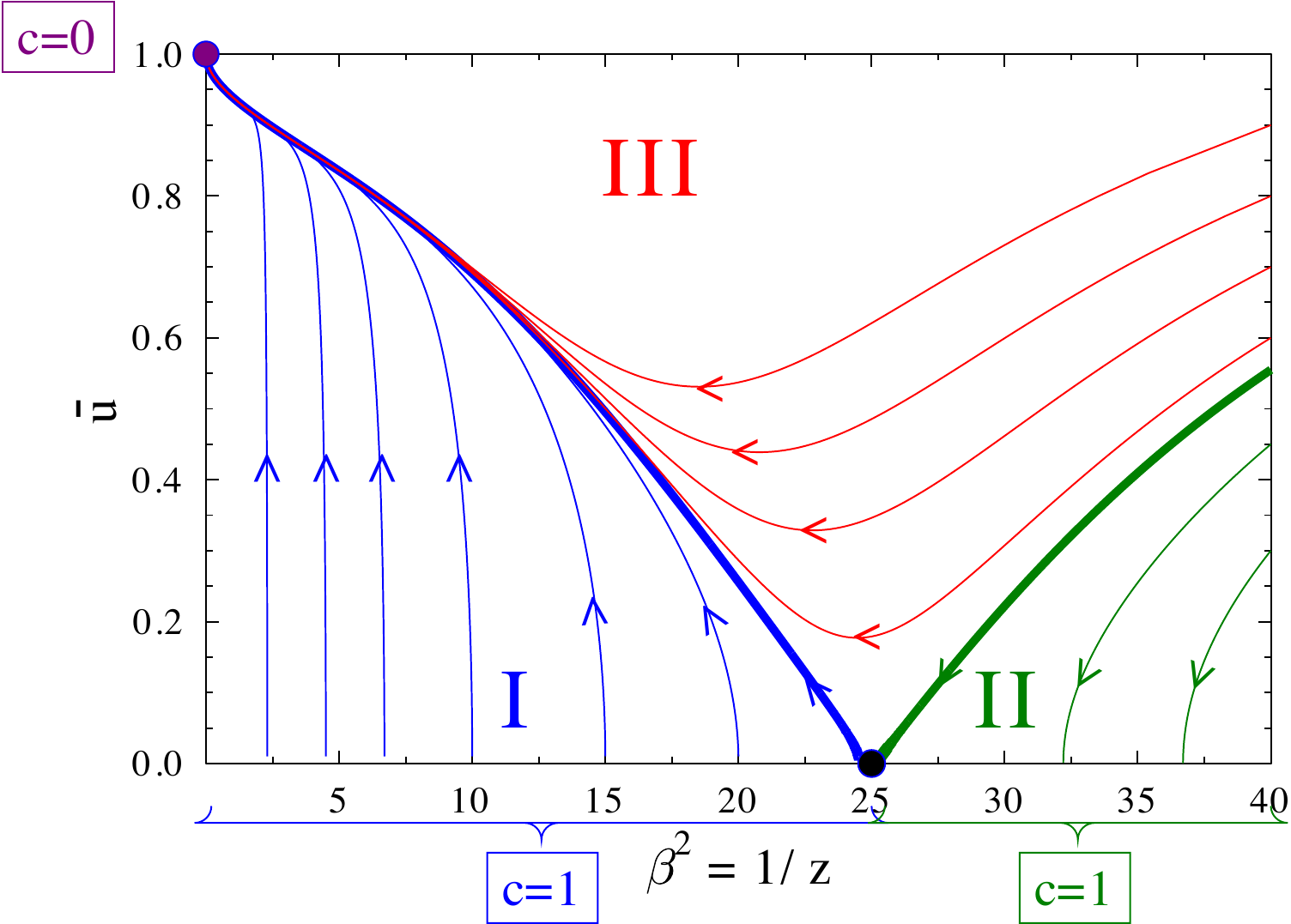}
\caption{
The flow diagram of the SG model in the scale-dependent frequency approximation. The phase space is 
divided into three regions. In region $I$ we have a line of UV repulsive Gaussian fixed points ($\bar u=0,
\beta^2<8\pi$). Every trajectory starting in the vicinity of this line ends in an IR attractive fixed point (purple 
full circle, $\bar u =1,\beta^2 =0$). The $\Delta c$ observed along the trajectories of this region should be 
equal to $1$. Region $II$ contains a line of IR attractive Gaussian fixed points ($\bar u =0,\beta^2 >8\pi,$) 
which are the end points of trajectories starting at $\beta^2 \approx\infty$ below the thick green line, i.e. the 
separatrix. Region $III$ contains those trajectories starting at $\beta^2 \approx\infty$ which end in the IR 
attractive fixed point (purple full circle).}
\label{fig2}
\end{center}
\end{figure}
The phase diagram obtained at this approximation level is sketched in Fig.\ref{fig2}, where we evidence three 
different regions. The $\Delta c$ is strictly well defined only in region $I$, where we start from a Gaussian fixed 
point $c_{UV}=1$ and we end up on a massive IR fixed point $c_{IR}=0$. The massive IR fixed point related 
to the degeneracy of the blocked action is an important feature of the exact RG flow \cite{IR1,IR2,IR3,IR4} and it was 
considered in SG type models \cite{SG_Trun,IR_SG1,IR_SG2}.

In region $II$ the trajectories end in the Gaussian fixed points $c=1$ but they are coming from infinity where 
actually no fixed point is present. This is due to the fact that we are not considering in our ansatz \eqref{eaans} 
any operator which can generate a fixed point at $c>1$ and then the trajectories ending at $c=1$ are forced to 
start at infinity. Thus, $\Delta c$ is not defined in this region.

Region $III$ contains those trajectories which start at $\beta=\infty$ but end in the IR massive fixed point at $c=0$. 
Even in this case the $\Delta c$ is not well defined.

In the following we are going to discuss in details the results of region $I$ where all the trajectories should give 
$\Delta c=1$. We shall ignore region $II$ where the $\Delta c$ is not defined, briefly discussing region $III$.

The presence of wave-function renormalization is necessary to obtain the qualitative correct flow diagram for the 
SG. Note that Eq. \eqref{c_function} has been derived only in the case of scale-independent kinetic term
and the derivation of an equivalent expression in the case of running wave function renormalization appears far more
demanding than the calculation sketched in this paper. However, it is still possible to get a sensible result
using the mapping between the running 
 wave-function renormalization and the running frequency $\beta_{k}$ cases (as shown in Eqs.~\eqref{Lpatransform} and \eqref{Lpatransform2}), finally obtaining Eq. \eqref{c_function_lpa_prime_rescaled} In other words the equation \eqref{c_function} is valid only at LPA level, but it is 
 still possible to apply it to the $z+$LPA scheme, since, thanks to the mapping described in Eqs.~\eqref{Lpatransform} and \eqref{Lpatransform2} the 
 $z+$LPA ansatz can be mapped into an LPA one. 

\textbf{We will then use directly the ansatz,}
\begin{equation}
\label{Ansatz_running_frequency}
\tilde{V}_{k}=\tilde{u}_{k}\cos(\beta_{k}\phi)
\end{equation}
with no wave function renormalization present in the kinetic term.
This ansatz is equivalent to ansatz \eqref{eaans} if we rescale the field and use the 
relations \eqref{Lpatransform} and \eqref{Lpatransform2}, with the running frequency playing
 the role of a wave-function renormalization. 

Ansatz \eqref{Ansatz_running_frequency} is not suited to study the SG model
when full periodicity has to be preserved, indeed when we substitute it into equation \eqref{ea_v}
symmetry breaking terms appear. The same happens
when we substitute it into Eq.\eqref{c_function}.
However in the latter case symmetry breaking terms are not dangerous,  since we have to
evaluate the expression at the potential minimum where all the symmetry breaking terms
vanish.

Proceeding in this way we obtain 
\begin{equation}
\label{c_function_lpa_prime}
k\partial_{k}c_{k}=\frac{(\beta_{k}^{2}k\partial_{k}\tilde{u}_{k}+2\tilde{u}_{k}\beta_{k}k\partial_{k}\beta_{k})^{2}}
{(1+\tilde{u}_{k}\beta_{k}^{2})^{3}}
\end{equation}
where no inconsistency is present.

 We still cannot use expression \eqref{c_function_lpa_prime}, since we cannot write
a flow for $\beta_{k}$ due to the non-periodic terms.
To avoid these difficulties we rewrite expression \eqref{c_function_lpa_prime} using the inverse transformation of 
\eqref{Lpatransform} and \eqref{Lpatransform2}, %
\begin{equation}
\label{c_function_lpa_prime_rescaled}
k\partial_{k}c_{k}=\frac{(k\partial_{k}\tilde{u}_{k})^{2}}
{(1+\tilde{u}_{k})^{3}}.
\end{equation}
The last expression is fully coherent and represents the flow of the $c$-function in presence of a running 
wave-function renormalization into the SG model; it is worth noting that the use of transformations \eqref{Lpatransform} 
and \eqref{Lpatransform2} gave us the possibility to derive the expression \eqref{c_function_lpa_prime} from Eq 
\eqref{c_function}, which was derived in \cite{c_func} in the case of no-wave-function renormalization.
%
%
\begin{figure}[ht] 
\begin{center} 
\includegraphics[width=0.48\textwidth]{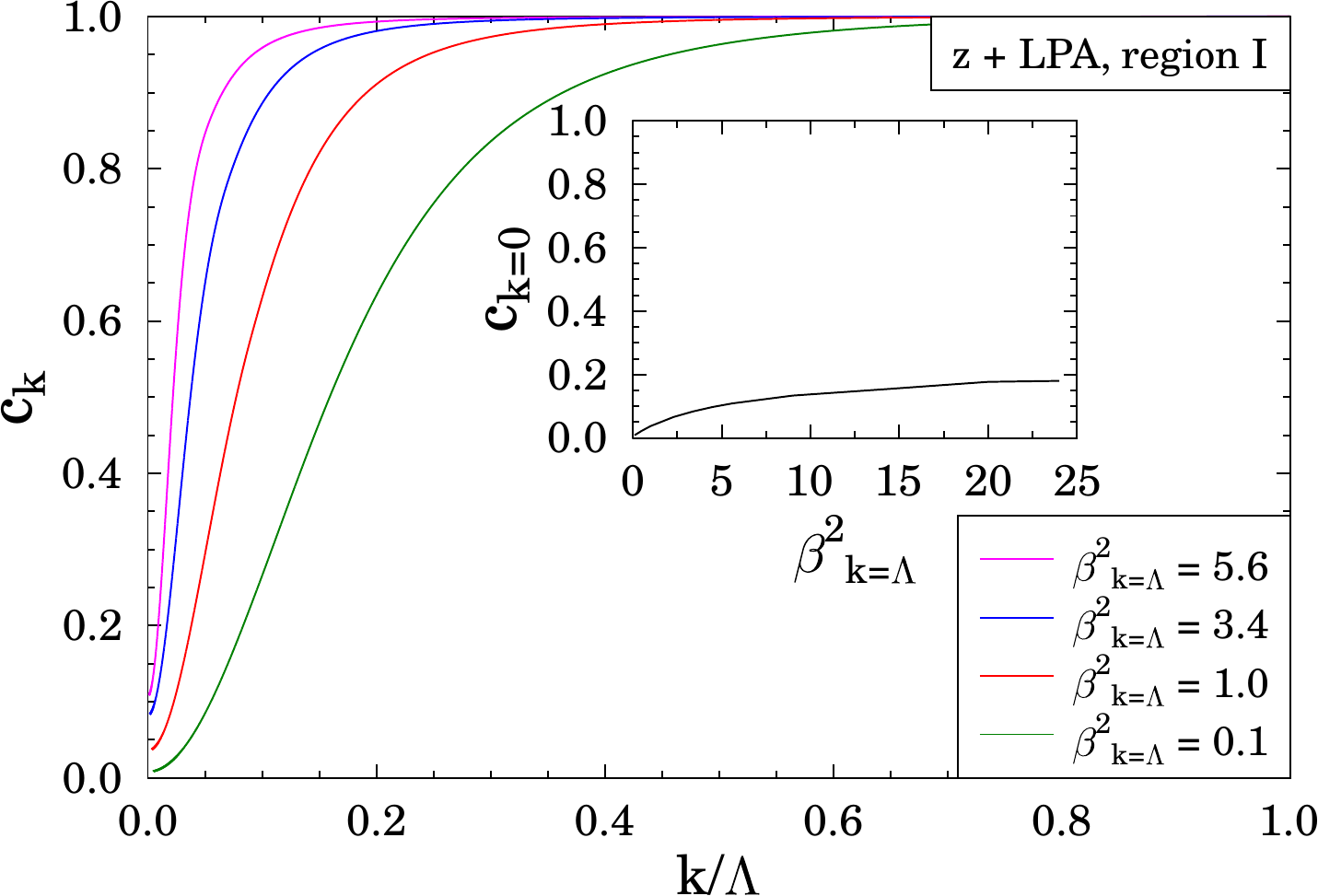}
\caption{
\label{fig3}
Running of the $c$-function obtained in the case of scale-dependent wave-function renormalization for the 
single-frequency SG model, as expected the case of small frequency ($\beta_{k=\Lambda}<0.1$) was already 
very well described by the scale-independent frequency case \fig{fig1}. The inset shows the results obtained for 
$c_{IR}$ as a function of $\beta_{k=\Lambda}$, these results show lower accuracy in the large frequency limit, 
while they become practically exact in the limit $\beta_{k=\Lambda}\to 0$, accordingly with \cite{c_func}.} 
\end{center}
\end{figure}

In the limit $\beta^2_{k=\Lambda} \to 0$, the IR result of the $c$-function (see the inset of \fig{fig3}) tends to 
zero. This implies that in the limit $\beta^2_{k=\Lambda} \to 0$ the difference $\Delta c \to 1$. The numerical 
result found in this case reaches the accuracy $1\geq \Delta c \geq 0.99$ of the scale-independent frequency solution 
\eq{sg_scale} but now  the periodicity of the SG model is fully preserved (which was not the case in \cite{c_func}). 
It should be also noted that the accurate results of Fig.\ref{fig3} could not be obtained in the mass cutoff framework 
\eqref{Morris_Family} with $b=1$, which does not allow the flow to converge, but our findings were obtained with 
the smoother $b=2$ cutoff. 

Fig \ref{fig3} reports the running of the $c$-function for various values of the initial condition $\beta_{\Lambda}$. 
The final $\Delta c$ value depends on the trajectory even if it should not be at exact level. This discrepancy shows that 
the flow obtained by approximated FRG procedure cannot satisfy the exact CFT requirements for the $c$-function.

The discrepancy between the exact $\Delta c=1$ value and the actual results obtained by the FRG approach can 
be used to quantify the error committed by the truncation ansatz in the description of the exact RG trajectories.

 We observe that the results of \fig{fig3} main and inset are obtained by using power-law regulator with $b=2$.
The same computation appears to be considerably more difficult using general cutoff functions, including the exponential one.
 
Let us note that for vanishing frequency the RG flow equations become regulator-independent and that the 
$c$-function value tends to the exact result $\Delta c=1$. This justifies the accuracy obtained in \cite{c_func} 
even though the mass cutoff was used and the periodicity violated.

Finally we go on showing the results in region $III$. As discussed in the description of \fig{fig2}, trajectories 
in region $III$ of the SG flow diagram should not have a well defined value for the $c$-function, due to the fact 
that those trajectories start at $\beta_{k=\Lambda}=\infty$ where no real fixed point is present.

\begin{figure}[!ht] 
\begin{center} 
\includegraphics[width=0.48\textwidth]{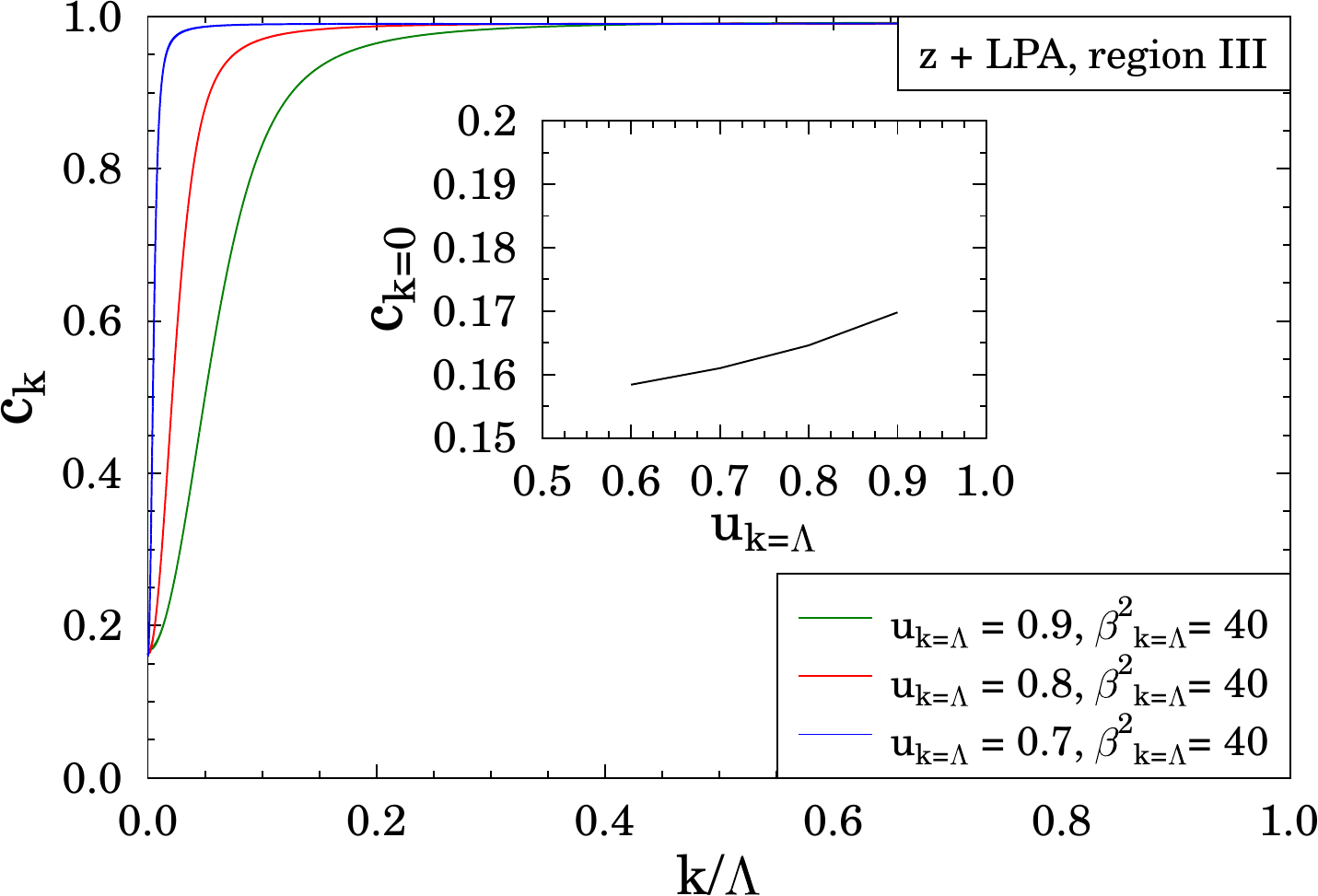}
\caption{The flow of the $c$-function in Region $III$ of the SG flow diagram \fig{fig2}, the result is approximately 
$\Delta c =1$ due to the fact that in region $III$ most of the contribution to the $c$-function comes from the part of 
the trajectories very close to "master trajectory" separatrix of region $I$ (the blue thick line in \fig{fig2}).
\label{fig4}} 
\end{center}
\end{figure}

However the numerical results obtained for those trajectories \fig{fig4} are not so far from $\Delta c=1$, due to the 
fact that they get most of the contribution in the region where they approach the "master trajectory" separatrix of region $I$ 
i.e. the blue thick line in Fig.\ref{fig2}, which we know to have a value $\Delta c\approx 1$, while the portion of the 
trajectories close to region $II$ get almost zero contribute.
%
%
The results of region $III$ are also in agreement with the findings of region $II$ (not shown) 
where $\Delta c\approx 0$ for all the initial conditions.

\section{Conclusions}
\label{sec_sum}
In this paper we provided an estimation of the $c$-function over the RG trajectories 
of the sine-Gordon (SG) model in the whole parameter space. Using this result we showed that the numerical 
functional RG study of 
the SG model with scale-dependent frequency recovers for $\beta^2<8\pi$ 
(region I of \fig{fig2}) the exact result $\Delta c=1$ with a good quantitative 
agreement while preserving the periodicity, 
which is the peculiar symmetry of the model. 
We also pointed out the dependence of this $c$-function calculation on the approximation level considered. For $\beta=0$ one retrieves 
directly $\Delta c=1$,  also in the scale-independent frequency case, while for $\beta \neq 0$,  again using scale 
independent frequency, we recover this result in the $\beta^2 \to 0$ limit, while increasing $\beta^2$ up to $8 \pi$ in region $I$ as a result 
of the used approximation the agreement becomes worst, remaining anyway reasonably good, as shown in \fig{fig1}.

Retrieving $\Delta c=1$ 
is the SG counterpart of the computation of $\Delta c$ for the sinh-Gordon model \cite{fring,dorey}. This result can be understood 
by noticing that the analytical continuation $\beta \to i \beta$ \cite{doreybook} may be expected not to alter the $\Delta c$ defined in the Zamolodchikov 
theorem, and that functional RG even in its crudest approximation does not spoil such correspondence for $\Delta c$, provided that the periodicity 
of the SG field is correctly taken into account.

We developed a fully coherent expression for the $c$-function in the case of running frequency, which gives better 
results in the whole region $I$ (defined in \fig{fig2}). These results are compatible with the exact scenario up to an accuracy of $80\%$ 
in the whole region $I$. Such accuracy grows to $99\%$ in the small beta region in agreement with \cite{c_func}, as we 
discussed in \fig{fig3}.

We also noticed that for $\beta^2 >0$ the use of the mass cutoff, 
as necessary to be consistent with expression \eqref{c_function}, is not possible due to bad convergence properties,
and the use of different $b$ values or of different cutoff types is needed.

It should be noted that while the numerical results are quite accurate the exact property 
that all trajectories of region $I$ 
should have the same value for $\Delta c$ is not preserved by truncations schemes \eqref{eaans} 
(results in \fig{fig1}) nor
\eqref{sg} (results in \fig{fig3}). Actually the $\beta^2\to 0$ 
limit always gives the correct result even when treated with the 
most rough truncation, 
this result being independent from the cutoff function. At variance 
one needs to go to the running frequency 
case to obtain reliable results for $\beta^2 >>0$.

 Even when full periodicity in the field is maintained, the $z+LPA$ truncation scheme is not sufficient to recover exact results for the $c$-function. Indeed the quantity computed using expression \eqref{c_function} satisfies two requirements
of Zamalodchikov's $c$-theorem,
\begin{enumerate}
\item $\partial_{t}c_{t}\geq 0$ along the flow lines,
\item $\partial_{t}c_{t}=0$ at the fixed points,
\end{enumerate}
 but it fails to reproduce the exact central charge of SG theory.

 The final result of our calculation also depends on the chosen cutoff function and,
as already mentioned, it was not possible to use the same cutoff scheme for both the couplings flow equations and the $c$-function flow \eqref{c_function}. 
We do not expect these issues to be responsible for the error in the fixed point value of the $c$-function. Modifications of the cutoff scheme in LPA calculations have small influence on the results (around 5\%) and we may expect this property to be maintained at $z+$LPA level, where calculations with different cutoff functions were not possible. 

The main source of deviation from the exact result $\Delta c=1$ is then probably due to $z+$LPA truncation in itself. We are not able to identify whether this deviation is only due to the approximation in the $c$-function flow or rather to the description of the fixed point given in $z+$LPA, which does not reproduce the exact central charge. 

Certainly the $c$-function flow at $z+$LPA level not merely violates the exact fixed point value of the $c$-function, but it is also  not able to produce trajectory independent results, as shown in \fig{fig3}. This scenario is not consistent with an unique central charge value at the SG fixed point and it is then impossible to dig out any information about this quantity from this approach. In this perspective it would be interesting to have an independent method to calculate the fixed point central charge at a given truncation level.

 Obviously the reproduction of the Zamalodchikov's $c$-theorem should be better satisfied increasing the truncation level considered. However it has been shown that, at LPA level, the addition of further harmonics in the potential does not improve the results presented, while the introduction of running frequency is crucial to achieve consistency of the phase diagram.

 This situation is peculiar of the LPA truncation level. Beyond $z+$LPA we expect the most relevant corrections from higher harmonics in the potential and only small variations are expected from the introduction of higher fields derivatives in \eqref{eaans}

Finally we remark that the trajectories of the other two regions do not have a definite $\Delta c$ value, however
while region $II$ gives results $\Delta c \approx 0$, region $III$ has the $\Delta c$ values close to the ones obtained
in region $I$ \fig{fig4}, due to the fact that all the trajectories in this region merge with the "master trajectory" separatrix of region 
$I$ in the $k\to 0$ limit.

We conclude by observing that in our opinion a relevant future extension of this work could be the study of the $c$-function 
for the Ising model and for minimal conformal models in general. Our work also points out to the possible investigation with RG techniques of 
the analytical continuation relating the sine- and the sinh-Gordon models. From this respect we think it could be worthwhile to systematically 
study models interpolating 
between these two celebrated cases, both to highlight the roaming phenomenon for integrable interpolations and to 
put forward critical properties of non-integrable interpolations.

\section*{Acknowledgement}

The authors gratefully thank A. Codello, G. Delfino,  B. Lima De Souza, G. Mussardo, I. G. M\'ari\'an,
R. Percacci and  G. Tak\'acs for useful discussions during the 
preparation of this work. This work was supported by the J\'anos Bolyai Research Scholarship of the 
Hungarian Academy of Sciences. Support from the CNR/MTA Italy-Hungary 2013-2015 Joint Project 
"Non-perturbative field theory and strongly correlated systems" is gratefully acknowledged. A.T. acknowledge support 
from the CNR project ABNANOTECH.


\begin{thebibliography}{99}

\bibitem{giuseppe}
G. Mussardo, {\em Statistical field theory: an introduction to exactly solved models 
in statistical physics} (Oxford, Oxford University Press, 2010).

\bibitem{Goldenfeld}
N. Goldenfeld, {\em Lectures on phase transitions and the renormalization group} 
(Reading, Addison Wesley, 1992). 

\bibitem{Cardy}
J. Cardy, {\em Scaling and renormalization in statistical physics} 
(Cambridge, Cambridge University Press, 1996). 

\bibitem{polyakov_1970}
A. M. Polyakov, JETP Lett. {\bf 12}, 381 (1970).

\bibitem{di_francesco_1997}
P. Di Francesco, P. Mathieu, and D. S\'en\'echal, {\em Conformal Field Theory} (New York, Springer, 1997).

\bibitem{c_theorem}
A. B. Zamolodchikov, JETP Lett. {\bf 43}, 730 (1986).

\bibitem{cardy_prl}
J. L. Cardy, Phys. Rev. Lett. {\bf 60}, 2709 (1988).

\bibitem{fring}
A. Fring, G. Mussardo, and P. Simonetti, Nucl. Phys. B {\bf 393}, 413 (1993).

\bibitem{dorey}
P. Dorey, G. Siviour, and G. Tak\'acs, JHEP {\bf 1503}, 54 (2015).

\bibitem{minnaghen}
P. Minnhagen, Rev. Mod. Phys. {\bf 59}, 1001 (1987).

\bibitem{kadanoff}
L. P. Kadanoff, {\em Statistical physics: statics, dynamics and renormalization} 
(Singapore, World Scientific, 2000). 

\bibitem{korepin}
V. E. Korepin, N. M. Bogoliubov, and A. G. Izergin, 
{\em Quantum Inverse Scattering Method and Correlation Functions} 
(Cambridge, Cambridge University Press, 1997).

\bibitem{tsvelik}
A. O. Gogolin, A. A. Nersesyan, and A. M. Tsvelik, {\em Bosonization and strongly correlated systems} 
(Cambridge, Cambridge University Press, 1998).

\bibitem{coleman}
S. Coleman, Phys. Rev. D {\bf 11}, 2088 (1975).

\bibitem{c_func}
A. Codello, G. D'Odorico, and C. Pagani, JHEP {\bf 1407}, 040 (2014).

\bibitem{Berges02} 
J. Berges, N. Tetradis, and C. Wetterich, Phys. Rep. {\bf 363}, 223 (2002).

\bibitem{Polony04} 
J. Polonyi, Central Eur. J. Phys. {\bf 1}, 1 (2004).

\bibitem{Delamotte12} 
B. Delamotte, in {\em Order, disorder and criticality: 
advanced problems of phase transition theory}, 
Yu. Holovatch ed. (Singapore, World Scientific, 2007)
[\verb|arXiv:cond-mat/0702365|].

\bibitem{css}
I. N\'andori, JHEP {\bf 1304}, 150 (2013).

\bibitem{Mo1994}  
T. R. Morris, Int. J. Mod. Phys. A {\bf 9}, 2411 (1994). 

\bibitem{opt_rg}  
D. F. Litim, Phys. Lett. B {\bf 486}, 92 (2000).

\bibitem{nandori_sg}
I. N\'andori, J. Polonyi, and K. Sailer, Phys. Rev. D {\bf 63}, 045022 (2001).

\bibitem{schemes}
I. N\'andori, S. Nagy, K. Sailer, and A. Trombettoni, Phys. Rev. D {\bf 80}, 025008 (2009). 

\bibitem{nandori_msg}
I. N\'andori, Phys. Rev. D {\bf 84}, 065024 (2011).

\bibitem{SG_Trun}
S. Nagy, I. N\'andori, J. Polonyi, and K. Sailer, Phys. Rev. Lett. {\bf 102}, 241603 (2009).




\bibitem{IR1}
N. Tetradis and C. Wetterich, Nucl. Phys. B {\bf 383}, 197 (1992).

\bibitem{IR2}
J. Braun, H. Gies, and D. D. Scherer, Phys. Rev. D {\bf 83}, 085012 (2011). 

\bibitem{IR3}
S. Nagy, Phys. Rev. D {\bf 86}, 085020 (2012).

\bibitem{IR4}
S. Nagy, J. Krizsan, and K. Sailer, JHEP {\bf 1207}, 102 (2012).

\bibitem{IR_SG1}
S. Nagy and K. Sailer, Int. J. Mod. Phys. A {\bf 28}, 1350130 (2013).

\bibitem{IR_SG2}
S. Nagy, Nucl. Phys. B {\bf 864}, 226 (2012); Annals Phys. {\bf 350}, 310 (2014).

\bibitem{doreybook}
P. Dorey, in {\em Conformal Field Theories and Integrable Models}, Z. Horv\'ath and L. Palla eds.,  
Lecture Notes in Physics Volume {\bf 498}, pg. 85 (Berlin, Springer-Verlag, 1997).

\end{thebibliography}
\end{document}